\documentstyle[12pt]{article}
\catcode`@=11
\newcount\@tempcntc
\def\@citex[#1]#2{\if@filesw\immediate\write\@auxout{\string\citation{#2}}\fi
  \@tempcnta\z@\@tempcntb\m@ne\def\@citea{}\@cite{\@for\@citeb:=#2\do
    {\@ifundefined
       {b@\@citeb}{\@citeo\@tempcntb\m@ne\@citea\def\@citea{,}{\bf ?}\@warning
       {Citation `\@citeb' on page \thepage \space undefined}}%
    {\setbox\z@\hbox{\global\@tempcntc0\csname b@\@citeb\endcsname\relax}%
     \ifnum\@tempcntc=\z@ \@citeo\@tempcntb\m@ne
       \@citea\def\@citea{,}\hbox{\csname b@\@citeb\endcsname}%
     \else
      \advance\@tempcntb\@ne
      \ifnum\@tempcntb=\@tempcntc
      \else\advance\@tempcntb\m@ne\@citeo
      \@tempcnta\@tempcntc\@tempcntb\@tempcntc\fi\fi}}\@citeo}{#1}}
\def\@citeo{\ifnum\@tempcnta>\@tempcntb\else\@citea\def\@citea{,}%
  \ifnum\@tempcnta=\@tempcntb\the\@tempcnta\else
   {\advance\@tempcnta\@ne\ifnum\@tempcnta=\@tempcntb \else \def\@citea{--}\fi
    \advance\@tempcnta\m@ne\the\@tempcnta\@citea\the\@tempcntb}\fi\fi}
\catcode`@=12

\def\barr{\begin{array}}
\def\earr{\end{array}}
\def\beq{\begin{equation}}
\def\eeq{\end{equation}}
\def\bea{\begin{eqnarray}}
\def\eea{\end{eqnarray}}
\def\bmath{\begin{displaymath}}
\def\emath{\end{displaymath}}
\def\bq{\begin{quote}}
\def\eq{\end{quote}}
\def\Re{\mbox{Re}}

\def\cF{{\cal F}}
\def\cL{{\cal L}}
\def\cM{{\cal M}}
\def\cO{{\cal O}}
\def\cT{{\cal T}}
\def\li{\lambda_i}
\def\lj{\lambda_j}

\def\apprle{\hspace{-0.1cm}\stackrel{\displaystyle <}{\sim}}

\def\slash#1{\setbox0=\hbox{$#1$}#1\hskip-\wd0\hbox to\wd0{\hss\sl/\/\hss}}
\def\snG{\mbox{{\footnotesize n}}_G}
\def\snR{\mbox{{\footnotesize n}}_R}
\def\nG{\mbox{n}_G}
\def\nR{\mbox{n}_R}

\def\g5{\gamma_5}
\def\lNN{\lambda_N}

\def\lN1{\lambda_{N_1}}

\def\mpla#1{{\em Mod.\ Phys.\ Lett.\ }{\bf A#1}}

\def\npb#1{{\em Nucl.\ Phys.\ }{\bf B#1}}
\def\plb#1{{\em Phys.\ Lett.\ }{\bf A#1}}
\def\plb#1{{\em Phys.\ Lett.\ }{\bf B#1}}
\def\prl#1{{\em Phys.\ Rev.\ Lett.\ }{\bf #1}}

\def\prd#1{{\em Phys.\ Rev.\ }{\bf D#1}}

\def\ptp#1{{\em Prog.\ Theor.\ Phys.\ }{\bf #1}}

\def\zpc#1{{\em Z.\ Phys.\ }{\bf C#1}}
\begin{document}

\begin{flushright}
RAL/94-113\\[-0.2cm]
June 1994
\end{flushright}
\vskip1cm

\begin{center}
{\bf{\large HEAVY-NEUTRINO EFFECTS ON {\boldmath $\tau$}-LEPTON
DECAYS}}\\[3cm]
{\large  A.~Pilaftsis}\footnote[1]{E-mail address:
pilaftsis@v2.rl.ac.uk }\\[0.3cm]
{\em Rutherford Appleton Laboratory, Chilton, Didcot, Oxon, OX11 0QX,
UK}
\end{center}
\vskip2cm
\centerline {\bf ABSTRACT}

\noindent
Minimal extensions of the Standard Model that are motivated
by grand unified theories or superstring models with an $E_6$
symmetry can naturally predict heavy neutrinos of Dirac or
Majorana nature. Such heavy neutral leptons violate the
decoupling theorem at the one-loop electroweak order and hence
offer a unique chance for possible lepton-flavour decays of the
$\tau$ lepton, {\em e.g.} $\tau \to eee$ or $\tau \to \mu\mu\mu$,
to be seen in LEP experiments. We analyze such decays in models
with three and four generations.

\newpage

Recently, it has been observed that the Standard Model (SM) with
more than one right-handed neutrino can dramatically
relax~\cite{BW,ZPC,HM} the suppression of heavy-light neutrino mixing
$s^{\nu_l}_L (\sim \sqrt{m_{\nu_l}/m_N})$ as
derived in usual ``see-saw" scenarios~\cite{YAN,CL}.
High Dirac mass terms are then allowed to be present in the theory
without contradicting low-energy constraints on the light neutrino
masses. As an immediate phenomenological consequence, it was
originally found that the one-loop vertex function relevant for the
lepton-flavour-violating decay of the Higgs boson~\cite{APetal}
shows a strong quadratic dependence of the heavy neutrino mass.
Such nondecoupling effects originating from heavy Majorana neutrino
masses have been taken into account in the leptonic flavour-changing
decays of the $Z$ boson, leading to rates that could be probed at
the CERN Large Electron Positron Collider (LEP)~\cite{KPS}.
A similar enhancement due to heavy neutrinos has recently been found
to take place in leptonic diagonal $Z$-boson
decays, yielding sizeable non-universality effects~\cite{BKPS}.

In this note we would like to analyze the phenomenological implications
of unified theories for the three-body decays of the $\tau$ lepton
into other three charged leptons, which we denote hereafter as
$l$, $l_1$, and $\bar{l}_2$. In fact, we find analytically that the
decay amplitude of $\tau \to ll_1\bar{l}_2$ increase quadratically with
the mass of the heavy Dirac or Majorana neutrino, which explicitly violates
the decoupling theorem~\cite{AC}.
In particular, we find quantitatively that the decays,
$\tau\to e^-e^-e^+$ and $\tau\to e^-\mu^-\mu^+$ [or the
complementary decays,
{\em i.e.}, $\tau\to \mu^-\mu^-\mu^+$ and $\tau\to \mu^-e^-e^+$],
deserve the biggest attention from the phenomenological point of view.
For completeness, we will present results for the decays $Z\to e\tau$
or $Z\to \mu\tau$, using updated constraints for lepton-violating mixing
angles.
Previous works on flavour-changing decays of the $Z$ boson
in a variety of models may be found in Refs.~\cite{LFstring,LFCNC}

In brief, we first outline the basic low-energy structure
of the two most popular extensions of the
SM that can naturally account for very light or strictly massless
neutrinos. The field content of these models is inspired by heterotic
superstring models~\cite{EW} or certain grand unified theories (GUTs)
based on the $SO(10)$ gauge group~\cite{WW}.
The low-energy limit of such theories can be realized in (i) the
SM with right-handed neutrinos~\cite{YAN,SV,ZPC} and (ii) the
SM with left-handed and right-handed neutral singlets~\cite{EW,WW,BSVMV}.
In addition, we will consider enhancements resulting from a possible
fourth sequential family of leptons and quarks.
Adopting now the notation of Ref.~\cite{ZPC}, the Yukawa sector of the SM
with a number $\nR$ of right-handed neutrinos,
$\nu^0_{Ri}$, in addition to $\nG$ left-handed ones, $\nu^0_{Li}$, reads
\beq
-\cL^\nu_Y\ =\ \frac{1}{2} (\bar{\nu}^0_L,\ \bar{\nu}^{0C}_R)\
M^\nu \left( \barr{c} \nu^{0C}_L\\ \nu^0_R \earr \right)\  +\ H.c.,
\eeq 
where the $(\nG+\nR) \times (\nG+\nR)$-dimensional  neutrino-mass matrix
$M^\nu$ is given by
\beq
M^\nu\ =\ \left( \barr{cc} 0 & m_D\\ m_D^T & m_M \earr \right)
\eeq 
Since $M^\nu$ is a complex symmetric matrix, it can always
be diagonalized by an $(\nG+\nR) \times (\nG+\nR)$ unitary matrix $U^\nu$
according to the common prescription: $U^{\nu T}M^\nu U^\nu =\hat{M}^\nu$.
We identify the first $\nG$ mass eigenstates, $\nu_i$,
with the known $\nG$ light neutrinos ({\it i.e.}, $\nG=3$),
while the remaining $\nR$ mass eigenstates, $N_j$,
are novel heavy Majorana neutrinos predicted by the model.
The quark sector of such an extension can completely be
described by the SM.
The couplings of the charged- and neutral-current interactions
are mediated by the mixing matrices $B$ and $C$, respectively.
For more details, the reader is referred to~\cite{ZPC}.
$B$ and $C$ are correspondingly $\nG\times (\nR +\nG)$- and
$(\nG +\nR)\times (\nG+\nR)$-dimensional matrices,
which are defined as
\beq
B_{lj}\ = \sum\limits_{k=1}^{\snG} V_{lk} U^{\nu\ast}_{kj}\quad
\mbox{and}\quad
C_{ij}\ =\ \sum\limits_{k=1}^{\snG}\ U^\nu_{ki}U^{\nu\ast}_{kj},
\eeq 
where $V$ is the leptonic Cabbibo-Kobayashi-Maskawa (CKM) matrix.
Note that the flavour-mixing matrices $B$ and $C$ satisfy a
number of identities that have been forced by the renormalizability of
the model~\cite{ZPC}.
With the help of these identities, one can derive useful relations between
mixings $B$, $C$ and heavy neutrino masses. For a model with two-right
handed neutrinos, for example, we obtain
\beq
B_{lN_1}\ =\ \frac{\rho^{1/4} s^{\nu_l}_L}{\sqrt{1+\rho^{1/2}}}\ , \qquad
B_{lN_2}\ =\ \frac{i s^{\nu_l}_L}{\sqrt{1+\rho^{1/2}}}\ ,
\eeq 
where $\rho=m^2_{N_2}/m^2_{N_1}$ is a mass ratio of the two heavy Majorana
neutrinos $N_1$ and $N_2$ that are predicted in such a model,
and $s^{\nu_l}_L$ is defined as~\cite{LL}
\beq
(s^{\nu_l}_L)^2 \ \equiv\  \sum\limits_{j=1}^{\snR} |B_{lN_j}|^2
\ \simeq\ \left( m_D^\dagger \frac{1}{m_M^2} m_D \right)_{ll}.
\eeq 
Furthermore, the mixings $C_{N_iN_j}$ can easily be obtained by
\bea
C_{N_1N_1} &=& \frac{\rho^{1/2}}{1+\rho^{1/2}}\ \sum\limits_{i=1}^{\snG}
(s^{\nu_i}_L)^2, \qquad C_{N_2N_2}\ =\ \frac{1}{1+\rho^{1/2}}\
\sum\limits_{i=1}^{\snG} (s^{\nu_i}_L)^2, \nonumber\\
C_{N_1N_2}&=& -C_{N_2N_1}\ =\ \frac{i\rho^{1/4}}{1+\rho^{1/2}}\
\sum\limits_{i=1}^{\snG} (s^{\nu_i}_L)^2.
\eea 
Our minimal scenario will then depend only on the masses
of the heavy Majorana neutrinos, $m_{N_1}$ and $m_{N_2}$ [or equivalently on
$m_{N_1}$ and $\rho$], and the mixing angles $(s^{\nu_i}_L)^2$,
which are directly constrained by low-energy data.

Another attractive scenario can be considered a superstring-inspired
extension of the SM, in which left-handed neutral singlets, $S_{Li}$,
in addition to the right-handed neutrinos, $\nu^0_{R_i}$, have been introduced.
In this scenario, the light neutrinos are strictly massless to all orders
of perturbation theory, if $\Delta L=2$ interactions are absent
from the model~\cite{WW}.
For the sake of simplicity, we will assume that the number of right-handed
neutrinos, $\nR$,  equals the number of the singlet fields $S_{Li}$.
After the spontaneous
break-down of the SM gauge symmetry, the Yukawa sector relevant for the
neutrino masses is given by~\cite{EW,WW}
\beq
-\cL^\nu_Y\ =\ \frac{1}{2} (\bar{\nu}^0_L,\ \bar{\nu}^{0C}_R,\ \bar{S}_L)\
\cM^\nu \left( \barr{c} \nu^{0C}_L\\ \nu^0_R \\ S^C_L \earr \right)\  +\ H.c.,
\eeq 
where the $(n_G+2\nR)\times (\nG+2\nR)$ neutrino-mass matrix takes the form
\beq
\cM^\nu \ \ =\ \ \left( \barr{ccc}
0 & m_D & 0 \\
m^T_D & 0 & M \\
0 & M^T & 0 \earr \right).
\eeq 
Since the neutrino matrix in Eq.~(8) has rank $2\nR$, this implies
that $\nG$ eigenvalues of $\cM^\nu$ will be zero. These $\nG$
massless eigenstates are identified with the ordinary light
neutrinos, $\nu_e$, $\nu_\mu$ and $\nu_\tau$~\cite{EW,WW}.
The remaining $2\nR$ Weyl fermions are degenerate in
pairs due to the fact that $L$ is conserved and so form $\nR$ heavy
Dirac neutrinos. A nice feature of the model is that the individual
leptonic quantum numbers may be violated~\cite{BSVMV,JWFV}.
The charged-current and neutral-current interactions
of the SM with left-handed and right-handed isosinglets can be found
in Ref.~\cite{BSVMV}.
To a good approximation, we assume that possible novel particles related to
the above unified theories, such as Pati-Salam leptoquarks or extra
charged and neutral gauge bosons are sufficiently heavy so as to
decouple completely from our low-energy processes.

Unified theories are constrained by a number of low-energy
experiments~\cite{LL}
and LEP data~\cite{BGKL}.
Experimental tests giving stringent constraints turn out to be
the neutrino counting at the $Z$ peak, the precise measurement of the
muon width $\mu \to e\nu_e\nu_\mu$, charged-current universality effects on
$\Gamma(\pi \to e\nu)/\Gamma(\pi\to \mu\nu)$, non-universality
effects on $B(\tau \to e \nu \nu)/B(\tau \to \mu \nu\nu)$, etc.
All these constraints, which are derived by the low-energy data mentioned
above, depend, more or less, on the gauge structure of the model under
consideration. In particular, interesting phenomenology could
arise from possible decays $Z\to N^*\nu$ at LEP, in case
$m_N\apprle M_Z$~\cite{PROD}.
For the present analysis, we consider that all heavy
neutrinos are much heavier than $M_Z$, and thus tolerating
the following upper limits~\cite{BGKL}:
\beq
(s^{\nu_e}_L)^2,\ (s^{\nu_\mu}_L)^2\ \ < \ \ 0.015, \quad
(s^{\nu_\tau}_L)^2\ \ < \ \ 0.070,\quad \mbox{and}\quad
(s^{\nu_\mu}_L)^2(s^{\nu_e}_L)^2 \ < \ \ 1.\ 10^{-8}.
\eeq 
The last constraint in Eq.~(9) comes from the non-observation of
the decay mode $\mu\to e\gamma$ or $\mu \to eee$.
Another limitation to the parameters of our model comes from the
requirement of the validity of perturbative unitarity that can be violated
in the limit of large heavy-neutrino masses.
A qualitative estimate for the latter may be obtained by requiring that
the total widths, $\Gamma_{N_i}$, and masses of neutrino fields $N_i$
satisfy the inequality $\Gamma_{N_i}/m_{N_i}< 1/2$.
In the limit of $m_{N_i} \gg M_W,\ M_Z,\ M_H$, the afore-mentioned
requirement leads to~\cite{ZPC}
\beq
\frac{\alpha_w}{4 M^2_W}\ m^2_{N_i}\ |C_{N_iN_i}|^2\ < \ 1/2,
\eeq 
with $\alpha_w=g^2_w/4\pi$.

Since Eq.~(9) tells us that either $(s^{\nu_e}_L)^2$ or
$(s^{\nu_\mu}_L)^2$ and not both of them could be as large as~0.01,
we will assume, for example, that $(s^{\nu_\mu}_L)^2\simeq 0$.
Consequently,  $B(\tau^- \to e^- e^- \mu^+)$, $B(\tau^-\to \mu^-\mu^- e^+)$
will be vanishingly small. There are then two possible decays
that are of potential interest, {\em i.e.},
\bea
 &\mbox{a}.& \ \tau^-\to e^- \mu^-\mu^+  ,  \nonumber\\
 &\mbox{b}.& \ \tau^- \to e^- e^- e^+ .
\eea 
Of course, one would equally assume that $(s^{\nu_e}_L)^2\simeq 0$
and $(s^{\nu_\mu}_L)^2\simeq 0.01$. In such a case, the complementary
decays where $e$ is replaced by muon in Eq.~(11) and vice versa will
be of interest. Furthermore, we have to stress the fact that
a simultaneous observation of $\tau\to eee$ and $\tau\to \mu\mu\mu$
cannot be compatible with experiments leading to the third inequality
of Eq.~(9).
The matrix element relevant for the decay
$\tau (p_\tau)\to l(p_l)l_1(p_1)\bar{l}_2(p_2)$
gets contributions from $\gamma$- and $Z$-mediated graphs that may be
found in Ref.~\cite{KPS} and box diagrams shown in Fig.~1.
These three transition elements are generically written down as follows:
\bea
  \cT_\gamma(\tau\to l l_1 \bar{l}_2)&=&
     -\frac{i\alpha_w^2s_w^2}{4M_W^2}
     \delta_{l_1 l_2} \bar{u}_{l_1}\gamma^\mu v_{l_2}\
   \bar{u}_{l}\Big[ F^{\tau l}_\gamma (\gamma_\mu-\frac{q_\mu\slash{q}}{q^2})
   (1-\g5)
    \nonumber\\
& &              -iG_\gamma^{\tau l} \sigma_{\mu\nu}\frac{q^\nu}{q^2}
                                (m_\tau(1+\g5)+m_{l}(1-\g5))\Big] u_\tau,
     \\
& & \nonumber\\
  \cT_Z(\tau\to l l_1 \bar{l}_2)&=&-\frac{i\alpha_w^2}{16M_W^2}\
   \delta_{l_1 l_2}
   F_Z^{\tau l} \bar{u}_{l}\gamma_\mu(1-\g5)u_\tau\ \bar{u}_{l_1}\gamma^\mu
   (1-4s_w^2-\g5)v_{l_2}
     \, \\
& & \nonumber\\
  \cT_{Box}(\tau \to l l_1 \bar{l}_2)&=&-\frac{i\alpha_w^2}{16M_W^2}\
  F_{Box}^{\tau ll_1l_2}\
   \bar{u}_{l}\gamma_\mu(1-\g5)u_\tau\ \bar{u}_{l_1}\gamma^\mu(1-\g5)v_{l_2}\ ,
\eea 
where $q=p_1+p_2$, $s^2_w=1-M^2_W/M^2_Z$, and with $\li=m^2_i/M^2_W$
(summation over light and heavy Majorana states with masses $m_i$
implied),
\bea
F_\gamma^{\tau l} &=& \sum_{i} B^\ast_{\tau i}B_{li}F_\gamma(\li)\ ,
            \\
G_\gamma^{\tau l} &=& \sum_{i} B^\ast_{\tau i}B_{li}G_\gamma(\li)\ ,
            \\
F_Z^{\tau l} &=& \sum_{ij} B^\ast_{\tau i}B_{lj}
     \Big[\delta_{ij}F_Z(\li)+C^\ast_{ij}G_Z(\li,\lj)+C_{ij}H_Z(\li,\lj)\Big],
            \\
F_{Box}^{\tau l l_1 l_2} &=&
  \sum_{ij} B^\ast_{\tau i} B^\ast_{l_2j}(B_{li}B_{l_1j}+B_{l_1i}B_{lj})
    \: F_{Box}(\li,\lj)
            \nonumber\\
  & & +\ \sum_{ij} B^\ast_{\tau i} B^\ast_{l_2i}B_{lj}B_{l_1j}\:
G_{Box}(\li,\lj)
\ ,
\eea 
are composite form factors that include multiplicative factors of
certain combinations of $B$ and $C$ matrices.
The photonic Inami-Lim form factors $F_\gamma$ and
$G_\gamma$~\cite{IL1}, as well as the form factors $F_Z$~\cite{IL1},
$H_Z$, $G_Z$, $F_{Box}$, and $G_{Box}$ are to be presented analytically in
Ref.~\cite{IP}.
It is, however, useful to discuss  the asymptotic limit of the composite
form factors described above in a model with two heavy Majorana neutrinos.
Using the expressions of Eqs.~(4) and (6) for the mixing matrices $B$ and $C$,
we find that for $\lN1 = m^2_{N_1}/M^2_W \gg 1$ and
$\rho=m^2_{N_2}/m^2_{N_1} \gg 1$,
\bea
F_\gamma^{\tau l} &\to & -\ \frac{1}{6}\; s_L^{\nu_\tau}s_L^{\nu_{l}}
\ln\lN1 ,   \\
G_\gamma^{\tau l} &\to & \frac{1}{2}\; s_L^{\nu_\tau}s_L^{\nu_{l}},
   \\
F_Z^{\tau l} &\to & -\; \frac{3}{2}s_L^{\nu_\tau}s_L^{\nu_{l}}\ln\lN1\;
               \nonumber\\
  && +\; s_L^{\nu_\tau}s_L^{\nu_{l}}\sum_{i=1}^{\snG}\ (s_L^{\nu_i})^2\;
                  \frac{\lN1 }{(1+\rho^{\frac{1}{2}})^2}\left(-\frac{3}{2}\rho
                   +\frac{-\rho+4\rho^{\frac{3}{2}}-\rho^2}
                         {4(1-\rho)}\ln\rho\right)\;
               ,\\
F_{Box}^{\tau ll_1l_2}&\to &-\; (s_L^{\nu_\tau}s_L^{\nu_{l}}\delta_{l_1l_2}
                           +s_L^{\nu_\tau}s_L^{\nu_{l_1}}\delta_{ll_2})
               \nonumber\\
            && +\; s_L^{\nu_\tau}s_L^{\nu_{l}}s_L^{\nu_{l_1}}s_L^{\nu_{l_2}}\;
               \frac{\lN1 }{(1+\rho^{\frac{1}{2}})^2}
               (-\rho-\frac{\rho+\rho^{\frac{3}{2}}+\rho^2}
                           {1-\rho}\ln\rho).
\eea 
In the limit $\rho \to 1$ and for $\lN1 \sim \lambda_{N_2} \sim \lNN \gg 1$,
Eqs.~(21) and~(22) take the form
\bea
F_Z^{\tau l}& \to &-\; \frac{3}{2}s_L^{\nu_\tau}s_L^{\nu_{l}}\ln\lNN\;
                -\; \frac{1}{2} s_L^{\nu_\tau}s_L^{\nu_{l}}\sum_{i=1}^{\snG}\
                             (s_L^{\nu_i})^2\lNN
               ,\\
F_{Box}^{\tau ll_1l_2}&\to &
               -\; (s_L^{\nu_\tau}s_L^{\nu_{l}}\delta_{l_1l_2}
                +s_L^{\nu_\tau}s_L^{\nu_{l_1}}\delta_{ll_2})\;
               +\; \frac{1}{2}s_L^{\nu_\tau}s_L^{\nu_{l}}s_L^{\nu_{l_1}}
                s_L^{\nu_{l_2}}\; \lNN.
\eea 
{}From Eqs.~(19)--(24), it is then obvious that all the one-loop functions,
$F_\gamma^{\tau l}$, $G_\gamma^{\tau l}$, $F_Z^{\tau l}$, and
$F_{Box}^{\tau ll_1l_2}$,
violate the decoupling theorem~\cite{AC}. Such a violation
is a common feature for all theories based on the spontaneous symmetry
breaking mechanism.
Taking the dominant nondecoupling parts of the
composite form factors into account, we arrive at the simple expression
for the branching ratios:
\bea
B(\tau^-\to e^- \mu^- \mu^+) &\simeq &
     \frac{\alpha_w^4}{24576\pi^3}\ \frac{m_\tau^4}{M_W^4}\
     \frac{m_\tau}{\Gamma_\tau}  \Big[ \ |F_{Box}^{\tau e \mu \mu}|^2
 + 2 (1-2s^2_w)\Re [F_Z^{\tau e}F_{Box}^{\tau e \mu\mu *}]\nonumber\\
&& +\ 8s^4_w|F_Z^{\tau e}|^2  \ \Big] \nonumber\\
&\simeq &      \frac{\alpha_w^4}{98304\pi^3}\ \frac{m_\tau^4}{M_W^4}\
\frac{m_\tau}{\Gamma_\tau}
\frac{m^4_N}{M^4_W}\ (s_L^{\nu_\tau})^2 (s_L^{\nu_{e}})^2
\Bigg\{ (s_L^{\nu_{\mu}})^4\nonumber\\
&& +\ 2(1-2s^2_w)(s_L^{\nu_\mu})^2 \sum_i (s_L^{\nu_i})^2
+ 8s^4_w \Big[\sum_i (s_L^{\nu_i})^2\Big]^2\ \Bigg\}.
\eea 
In the same heavy neutrino limit, we obtain
\bea
B(\tau^-\rightarrow e^- e^- e^+) & \simeq &
     \frac{\alpha_w^4}{24576\pi^3}\ \frac{m_\tau^4}{M_W^4}\
     \frac{m_\tau}{\Gamma_\tau} \Big[ \frac{1}{2}|F_{Box}^{\tau eee}|^2
+2(1-2s^2_w)\Re [F_Z^{\tau e}F_{Box}^{\tau eee *}] \nonumber\\
&& +\ 12 s^4_w|F_Z^{\tau e}|^2 \Big] \nonumber\\
&\simeq &      \frac{\alpha_w^4}{98304\pi^3}\ \frac{m_\tau^4}{M_W^4}\
\frac{m_\tau}{\Gamma_\tau}
\frac{m^4_N}{M^4_W}\ (s_L^{\nu_\tau})^2 (s_L^{\nu_e})^2
\Bigg\{ \frac{1}{2}(s_L^{\nu_e})^4\nonumber\\
&& +\ 2(1-2s^2_w)(s_L^{\nu_e})^2  \sum_i (s_L^{\nu_i})^2
+ 12s^4_w \Big[\sum_i (s_L^{\nu_i})^2\Big]^2\ \Bigg\}.
\eea
In Eqs.~(25) and~(26), $\Gamma_\tau$ denotes the total decay width
of the $\tau$ lepton, which is experimentally measured to be
$\Gamma_\tau=2.16\ 10^{-12}$~GeV~\cite{PDG}.

Apart from the $\tau$-lepton decays given in Eq.~(11), the decay
$Z\to e\tau$ can also be enhanced due to the same heavy neutrino
effects to an extend that may be seen at LEP~\cite{KPS}.
To the leading order of heavy neutrino masses ($m_N \gg M_W$),
the branching ratio of this decay mode is obtained by
\bea
B(Z\to \tau^- e^+  + e^- \tau^+) &=&
\frac{\alpha_w^3}{48\pi^2c_w^3}\frac{M_W}{\Gamma_Z}
                              |\cF_Z^{e\tau}(M^2_Z)|^2 \nonumber\\
&\simeq & \frac{\alpha_w^3}{768\pi^2c_w^3}\frac{M_W}{\Gamma_Z}
\frac{m^4_N}{M^4_W} (s_L^{\nu_e})^2 (s_L^{\nu_\tau})^2
\Big[ \sum_i (s_L^{\nu_i})^2 \Big]^2,
\eea 
where $\Gamma_Z$ is the total width of the $Z$ boson. Note that
$\cF_Z^{e\tau}(0)= F_Z^{e\tau}/2$.

In order to minimize the free parameters of the theory that
could vary independently,
we will assume an extension of the SM by two
right-handed neutrinos. The neutrino mass spectrum of such a model
consists of three light Majorana neutrinos which have been identified
with the three known neutrinos, $\nu_e$, $\nu_\mu$, and $\nu_\tau$, and
two heavy ones denoted by $N_1$ and $N_2$. On the other hand,
the SM inspired by superstring theories with an $E_6$ symmetry~\cite{EW}, in
which one left-handed and one right-handed chiral singlets are present,
can effectively be recovered by the SM with two right-handed neutrinos when
going to the degenerate mass limit for the two heavy Majorana neutrinos.

Assuming the maximally allowed values~\cite{BGKL}
for $(s^{\nu_\tau}_L)^2=0.07$ and
$(s^{\nu_e}_L)^2=0.015$ $((s^{\nu_\mu}_L)^2\simeq 0)$ given in Eq.~(9),
we find the encouraging branching ratios
\beq
B(\tau^- \to e^- e^- e^+) \ \apprle\ 2.\ 10^{-6}\quad \mbox{and}
\quad
B(\tau^- \to e^- \mu^- \mu^+)\ \apprle\ 1.\ 10^{-6},
\eeq 
where the upper bounds is estimated by using $m_N\simeq 3$~TeV as derived
from Eq.~(10).
The present experimental upper limits on these decays are~\cite{PDG}
\beq
B(\tau^- \to e^- e^- e^+),\ B(\tau^- \to e^- \mu^- \mu^+)\ \ < \ \
1.4\ 10^{-5},\quad \mbox{CL}=90\%.
\eeq 
Even if we assume smaller values for the mixing angles,
$(s^{\nu_\tau}_L)^2=0.035$ and $(s^{\nu_e}_L)^2=0.01$ $((s^{\nu_\mu}_L)^2=0)$,
the lepton-flavour-violating decays of the $\tau$ lepton can still be
significant, {\em i.e.},
\beq
B(\tau^- \to e^- e^- e^+) \ \apprle\ 5.\ 10^{-7}\quad \mbox{and}
\quad
B(\tau^- \to e^- \mu^- \mu^+)\ \apprle\ 3.\ 10^{-7}.
\eeq 
Since the branching ratio increase with the heavy neutrino mass
to the fourth power, this strong mass dependence gives rise to measurable
values for the leptonic three-body decays of the $\tau$ lepton.
To be precise, if we had neglected contributions of
seemingly suppressed terms $\sim (s^{\nu_i}_L)^4$ in the transition
amplitude,  we would then have found a reduction
of our numerical values up to $\sim 10^{-2}$. In the low-mass range of
heavy neutrinos ({\it i.e.}~for $m_{N_i}< 200$~GeV) the difference between
the complete and the approximate computation
is quite small and consistent with results
obtained in~\cite{JWFV}. For very heavy neutrinos,
the situation is quite different, since in
the decay amplitude, terms proportional
to $(s^{\nu_i}_L)^2$ increase logarithmically with the heavy neutrino
mass $m_N$, {\it i.e.}~$\ln(m_N^2/M^2_W)$, while terms of
$\cO ((s^{\nu_i}_L)^4)$ show a strong quadratic dependence in the heavy
neutrino mass, i.e.~$m_N^2/M^2_W$.
Finally, $\tau$ leptons can also decay hadronically via the channels:
$\tau \to l_i\eta$, $\tau\to l_i\pi^0$, etc.~\cite{JWFV}.
Since present experimental sensitivity to these decays is
rather weak~\cite{PDG}, {\em e.g.},
$B(\tau\to e\pi^0)<1.4\ 10^{-4}$, at CL$=90\%$, one could
expect that it would be difficult to probe
heavy neutrino effects in such hadronic decay channels.

We will now investigate the LEP potential of observing
lepton-flavour-violating decays at the $Z$ peak.
Since we always assume that $(s^{\nu_\mu}_L)^2\simeq 0$ for reasons
mentioned above, we will focus our
analysis on the decays $Z\to e^-\tau^+$ or $e^+\tau^-$.
Within the perturbatively allowed range of heavy neutrino masses, we find
\bea
B(Z\to e^-\tau^+\ +\ e^+\tau^-)\ \apprle \ 4.0\ 10^{-6}, \quad\mbox{for}
\quad (s^{\nu_\tau}_L)^2=0.070, \quad (s^{\nu_e}_L)^2=0.015,\nonumber\\
B(Z\to e^-\tau^+\ +\ e^+\tau^-)\ \apprle \ 1.1\ 10^{-6}, \quad\mbox{for}
\quad (s^{\nu_\tau}_L)^2=0.035, \quad (s^{\nu_e}_L)^2=0.010,\nonumber\\
B(Z\to e^-\tau^+\ +\ e^+\tau^-)\ \apprle \ 6.0\ 10^{-7}, \quad\mbox{for}
\quad (s^{\nu_\tau}_L)^2=0.020, \quad (s^{\nu_e}_L)^2=0.010.
\eea 
All these branching ratios could be detected at
future LEP data, as the present experimental sensitivity at LEP
is~\cite{PDG}
\beq
B(Z\to e^-\tau^+\ +\ e^+\tau^-)\ < 1.3\ 10^{-5}, \quad \mbox{CL}=95\%.
\eeq 
However, the present upper bound on the flavour-changing
$Z$-boson decays do not yet impose any severe constraints on our
analysis of $\tau$-lepton decays.

In the following, we will briefly discuss the phenomenological
consequences induced by the presence of an extra sequential family.
Such scenarios have recently received much attention due to the
additional fact that they could naturally resuscitate extended
technicolour theories~\cite{ETC}. LEP precision experiments
provide a useful framework to either constrain or establish such extended
models, when one analyzes electroweak oblique parameters~\cite{BS,STU,WVX}
and other quantum effects~\cite{Bernd}. As a consequence of such an analysis,
the lightest of the two heavy Majorana neutrinos belonging to the
fourth family cannot be heavier than~1~TeV. In our models, the inclusion
of an extra family amounts to replacing $\sum_i (s^{\nu_i}_L)^2 \to 1$ in
Eqs.~(25), (26), and~(27). The branching-ratio values we obtain in such
models are larger, {\em i.e.},
\bea
B(\tau^- \to e^- e^- e^+),\ B(\tau^- \to e^- \mu^- \mu^+)
\ \apprle\ 4.\ 10^{-6}, && \hspace{2cm}\\
B(Z\to e^-\tau^+\ +\ e^+\tau^-)\ \apprle \ 8.\ 10^{-6}. &&
\eea

In conclusion, we have explicitly demonstrated that
GUT or superstring-inspired extensions of the minimal SM
can naturally account for  sizeable branching ratios of the
$\tau$-lepton decays of the type $\tau\to eee$,
$\tau\to \mu\mu\mu$, etc. These decays  show a strong {\em quadric}
mass dependence of the heavy neutrino mass (see Eqs.~(25) and (26)),
which gives a unique chance for such non-SM signals to be seen in
present or future $\tau$ factories.\\[0.8cm]
{\bf Acknowledgements.} I wish to thank Amon Ilakovac, Bernd Kniehl,
and Jose Valle for useful discussions and comments.


\newpage

\centerline{\bf\Large Figure Caption }
\vspace{-0.2cm}
\newcounter{fig}
\begin{list}{\bf\rm Fig. \arabic{fig}: }{\usecounter{fig}
\labelwidth1.6cm \leftmargin2.5cm \labelsep0.4cm \itemsep0ex plus0.2ex }

\item Feynman diagrams relevant for the leptonic decays
$\tau \to l l_1 \bar{l}_2$.

\end{list}

\end{document}